\documentclass[lettersize,journal]{IEEEtran}
\usepackage{amsmath,amsfonts}
\usepackage{array}
\usepackage{textcomp}
\usepackage{stfloats}
\usepackage{url}
\usepackage{verbatim}
\usepackage{graphicx}
\usepackage{multirow}
\usepackage{makecell}
\usepackage{float}
\usepackage{hyperref}
\usepackage{subfigure}
\usepackage{algorithm}
\usepackage{booktabs}
\usepackage{algorithmicx}
\usepackage{algpseudocode}

\begin{document}

\title{Advancing CAN Network Security through RBM-Based Synthetic Attack Data Generation for Intrusion Detection Systems}

\author{Huacheng Li, Jingyong Su, Kai Wang

\thanks{This work is supported by National Natural Science Foundation of China (NSFC) (grant number 62272129) and Taishan Scholar Foundation of Shandong Province (grant number tsqn202408112).}

\thanks{Huacheng Li and Jingyong Su are with School of Computer Science and Technology, Harbin Institute of Technology, Shenzhen, China (e-mail: lihuacheng@stu.hit.edu.cn; sujingyong@hit.edu.cn).}

\thanks{Kai Wang is with School of Computer Science and Technology, Harbin Institute of Technology, Weihai, China, and also with Shandong Key Laboratory of Industrial Network Security, China (e-mail: dr.wangkai@hit.edu.cn).}

\thanks{Corresponding author: Kai Wang.}
}

\markboth{Journal of \LaTeX\ Class Files,~Vol.~14, No.~8, August~2021}%
{Shell \MakeLowercase{\textit{et al.}}: A Sample Article Using IEEEtran.cls for IEEE Journals}

\IEEEpubid{0000--0000/00\$00.00~\copyright~2021 IEEE}

\maketitle

\begin{abstract}
The rapid development of network technologies and industrial intelligence has augmented the connectivity and intelligence within the automotive industry. Notably, in the Internet of Vehicles (IoV), the Controller Area Network (CAN), which is crucial for the communication of electronic control units but lacks inbuilt security measures, has become extremely vulnerable to severe cybersecurity threats. Meanwhile, the efficacy of Intrusion Detection Systems (IDS) is hampered by the scarcity of sufficient attack data for robust model training. To overcome this limitation, we introduce a novel methodology leveraging the Restricted Boltzmann Machine (RBM) to generate synthetic CAN attack data, thereby producing training datasets with a more balanced sample distribution. Specifically, we design a CAN Data Processing Module for transforming raw CAN data into an RBM-trainable format, and a Negative Sample Generation Module to generate data reflecting the distribution of CAN data frames denoting network intrusions. Experimental results show the generated data significantly improves IDS performance, with CANet accuracy rising from 0.6477 to 0.9725 and EfficientNet from 0.1067 to 0.1555. Code is available at \url{https://github.com/wangkai-tech23/CANDataSynthetic}.
\end{abstract}

\begin{IEEEkeywords}
Internet of Vehicles, Controller Area Network, Synthetic Data Generation,  Restricted Boltzmann Machine.
\end{IEEEkeywords}

\section{Introduction}
\IEEEPARstart{I}{n} the era of Internet of Vehicles (IoV), the increasing connectivity and intelligence of vehicles have brought about numerous conveniences, but also exposed them to severe cybersecurity threats. The Controller Area Network (CAN), as a key communication protocol in vehicles, lacks inherent security measures, making it vulnerable to various cyber-attacks such as DoS attacks, fuzzy attacks, spoofing attacks, etc. \cite{Koscher2010ExperimentalSA,Ghane2020PreservingPI,Farivar2019DetectionAC,Lin2012CyberSecurityFT}. These attacks can disrupt vehicle control systems and even lead to catastrophic failures, endangering the safety of passengers and the normal operation of the transportation system.

To address these security issues, intrusion detection systems (IDS) play a crucial role. However, traditional IDS face significant challenges in training due to the scarcity of sufficient CAN attack data and the imbalance between normal and attack data samples. Although some machine learning techniques have been applied to CAN network security, such as structured anomaly detection and the use of long short-term memory (LSTM) networks \cite{Mter2010ASA,Taylor2016AnomalyDI,Wu2020ASO}, the problem of data shortage still persists.

Fortunately, generative models like Generative Adversarial Networks (GAN) and Diffusion models have emerged as potential solutions to enhance the performance of various models for classification tasks. However, the training process of GAN is prone to mode collapse, which severely affects the quality and diversity of the generated data. Meanwhile, Diffusion model typically requires a long training period and is more adept at learning the distribution of image data. However, it struggles to capture the distribution of data with explicit structures such as data frames and tables, which possess discrete semantic representations and intricate dependency relationships. This shortcoming restricts its application in generating CAN-related data.

\IEEEpubidadjcol

The Restricted Boltzmann Machine (RBM), as an energy-based model, is guided by the Contrastive Divergence (CD) algorithm to learn the probability distribution of the input data through an optimization process that minimizes the system's energy state.
RBM's proficiency in modeling the distribution of semantically complex data render it a preferable option for generating data that closely resembles the statistical properties of CAN traffic. 

In this paper, we designed a CAN Data Processing Module to transform raw CAN data into a format suitable for RBM training. Moreover, a Negative Sample Generation Module is devised to generate data that reflects the complex data distribution of CAN traffic, thereby producing more negative samples to balance the training dataset of the network intrusion detection model.

%\IEEEpubidadjcol

The primary contributions of this work are as follows:
\begin{itemize}
\item{We conduct an in-depth analysis of the CAN protocol and its associated security threats, providing valuable insights for IoV security research.}
\item{We propose an effective data generation framework based on RBM to address the issue of data imbalance in CAN network security research.}
\item{Through extensive experiments, we demonstrate that the generated data significantly improves the performance of existing IDS, validating the effectiveness of our approach.}
\end{itemize}

The rest of this paper is structured as follows: Section II examines the related work regarding vehicular network security and data generation techniques. Section III first presents the structure and characteristics of CAN data frames, then introduces the cybersecurity threats faced by CAN, and finally describes the structure of the RBM and its application in data generation. Section IV describes our methodology for generating synthetic attack data for CAN networks. Section V elaborates on the experimental setup and results. Section VI assesses the performance of the proposed approach. Finally, Section VII summarizes the paper and proposes future research directions.

\section{Related Work}

\subsection{Intrusion Detection}
Prevailing intrusion detection methods in the realm of in-vehicle networks can be categorized as follows:

\textbf{Signature-based IDS:} These systems rely on predefined signatures of known attacks and detect an intrusion when an event matches a stored signature. Their main advantage is the high speed of detecting known attacks. However, they fail to detect novel attacks as they cannot recognize patterns not in their database. In the CAN context, any variation in attack patterns can lead to undetected threats \cite{Wu2020ASO,Cho2016ErrorHO}.

\textbf{Anomaly-based IDS}: Such IDS constructs a profile of the normal network traffic behavior. Deviations from the predefined normal profile are flagged as potential intrusions, endowing it with the potential to detect novel attacks. However, its performance highly depends on training data quality and quantity. Taylor et al. \cite{Taylor2016AnomalyDI} employed long short-term memory (LSTM) networks for anomaly detection. Still, the data problem persisted as a challenge. 

\textbf{Time Interval-based IDS}: These IDS exploit the periodic characteristic of CAN transmissions. Anomalies can be detected by measuring the time intervals among messages. Müter and Asaj \cite{Mter2011EntropybasedAD} investigated an entropy-based method based on this mechanism. However, they struggle with sporadic transmissions and lack accurate methods for threshold determination and comprehensive evaluation.  

\textbf{Sequence-based IDS}: This type of IDS uses sequences of message identifiers (AIDs) to model the normal network state and is highly applicable across CAN-based IVNs due to the ease of compiling such sequences. However, interpreting abnormal sequence classifications can be difficult and replay attacks that mimic legitimate sequences may mislead the system. Marchetti and Stabili \cite{Marchetti2017AnomalyDO} constructed a transition matrix for adjacent AIDs, nevertheless, they encountered difficulties in optimizing the sequence length and minimizing false positives. 

\textbf{Payload-based IDS}: These systems focus on evaluating CAN message payloads in bit sequences. Although potentially effective, they require careful training set preparation to avoid false alarms from payload dynamics. Kang and Kang \cite{Kang2016IntrusionDS} suggested a binary intrusion detection method using 64 - dimensional bit sequences. However, its evaluation with limited streams and potentially unrealistic simulated payloads restricted practical use. Taylor et al. \cite{Taylor2015FrequencybasedAD} also had issues regarding window size selection and false positives when analyzing payloads via sliding window.

\subsection{Data Generation}
The application of generative models to expand training datasets has captured significant attention, emerging as a promising avenue in contemporary research.  

\textbf{Generative Adversarial Networks (GAN)}: GAN has demonstrated significant potential in generating data for diverse applications. Seo et al. \cite{Seo2018GIDSGB} pioneered GAN for automotive CAN network intrusion detection. However, prior GAN-based approaches often had coarse CAN message block construction, failing to accurately represent actual blocks during detection and struggling with data tampering threats. Xie et al. \cite{Xie2021ThreatAF} enhanced performance by integrating the CAN communication matrix. 

\textbf{Variational Autoencoders (VAE)}: Comprising an encoder and a decoder, VAEs form another class of generative models that can generate data mirroring the original distribution. In CAN security, despite being less explored than GANs, they have the potential to yield diverse and realistic synthetic data. However, akin to GANs, they struggle to precisely capture the complex inherent dependencies and structures of CAN data. 

\textbf{Diffusion Models}: Comprising forward and reverse diffusion processes, Diffusion Models have shown outstanding skills in generating image data. However, for CAN data generation, they struggle due to discrete semantic representations and complex dependencies in CAN data frames and tables. Additionally, their relatively long training times pose a significant hurdle for real-time automotive applications.

\section{Background}
\subsection{CAN Protocal}
CAN protocol is a robust communication protocol initially developed for automotive applications, facilitating reliable data exchange between ECUs within a vehicle. It has since been adopted in various industrial settings due to its high reliability, error detection capabilities, and ability to operate in noisy environments. Within the CAN protocol, data frames play a crucial role in the transmission of information. As shown in Figure \ref{fig:The CAN frame in Standard Format}, a data frame is structured to include several segments: Start of Frame (SOF), Arbitration Field, Control Field, Data Field, CRC Field, ACK Field, and End of Frame (EOF). This structured format ensures that messages are prioritized, identified, and transmitted with error checking measures in place. The meanings of various fields of the CAN data frame are described below:

\begin{figure}
    \centering
    \includegraphics[width=1.0\linewidth]{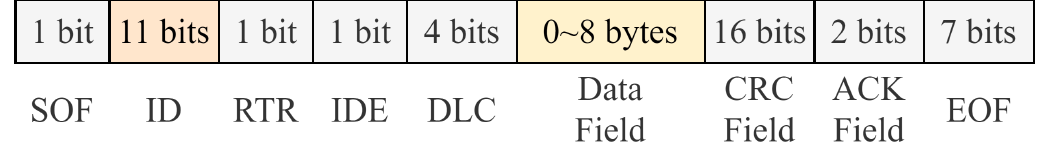}
    \caption{Demonstration of the CAN frame in Standard Format.}
    \label{fig:The CAN frame in Standard Format}
\end{figure}

\begin{itemize}
\item{Start of Frame (SOF): The data frame begins with a single dominant `0' bit, signaling the start of a new frame.}
\item{Identifier (ID): A unique identifier for the message, which can be either 11 bits (standard format) or 29 bits (extended format). The lower the ID value, the higher the priority of the message.}
\item{Remote Transmission Request (RTR): A single bit that differentiates between data frames and remote frames. A data frame (RTR = 0) carries actual data, while a remote frame (RTR = 1) requests data from another node.}
\item{Identifier Extension (IDE): A bit that indicates whether the frame is using a standard or extended format.}
\item{Data Length Code (DLC): A 4-bit field that specifies the number of data bytes that follow in the Data Field.}
\item{Data Field: This is a variable-length section that carries the actual data payload, ranging from 0 to 8 bytes.}
\item{Cyclic Redundancy Check (CRC) Field: A 16-bit CRC field is used to detect errors in the frame. It is calculated based on the ID and data bytes to ensure data integrity.}
\item{Acknowledge (ACK) Field: It includes an ACK slot and an ACK delimiter. The ACK slot is where the receiving node acknowledges receipt of the frame by transmitting a dominant `0' bit back to the sender.}
\item{End of Frame (EOF): The frame ends with seven consecutive recessive `1' bits, signaling the end of the frame.}
\end{itemize}

\subsection{Messages Injection Attack}
Within the CAN, message injection attacks represent a significant threat vector, leveraging vulnerabilities to disrupt or manipulate vehicle control systems. We introduce the following three general categories of attacks.
\subsubsection{DoS Attack}
DoS attacks are insidious in their simplicity yet devastating in their impact. They aim to incapacitate the CAN bus by flooding it with superfluous traffic, effectively denying legitimate nodes access to the network. This can be executed by continuously transmitting messages without respite, thereby monopolizing the bus and precluding other nodes from communication.

\subsubsection{Fuzzy Attack}
Fuzzy attacks exploit the inherent ambiguities in message interpretation and prioritization mechanisms of the CAN protocol. These attacks inject specially crafted messages into the network, designed to either confuse legitimate nodes or lead to unpredictable system behavior. Fuzzy attacks often involve priority manipulation, where attackers use misleading identifiers to alter the arbitration process, potentially causing critical messages to be delayed or ignored.

\subsubsection{Spoofing Attack}
Spoofing attacks involve the malicious impersonation of legitimate nodes or messages, with the intent to deceive the system and gain unauthorized access or manipulate control units. This form of attack is executed through tactics such as identity spoofing, where attackers broadcast messages with identifiers mimicking trusted nodes, leading other network participants to accept and act upon these fraudulent messages. Message spoofing involves the crafting and transmission of false data frames, sowing confusion among legitimate nodes. Furthermore, a man-in-the-middle (MitM) attack can undermine communication integrity by intercepting and potentially altering messages.

\subsection{Boltzmann Distribution}
In the realm of statistical mechanics, the Boltzmann distribution, a fundamental concept, describes the probabilistic distribution of particles across different energy states within a system and quantifies the likelihood of particles occupying a specific energy state based on the system's energy and temperature. The Boltzmann distribution is given by Equation \eqref{keq1}.

\begin{align}\label{keq1}
P_\alpha = \frac{1}{Z}exp(-\frac{E_\alpha}{kT})
\end{align}

$P_\alpha$ denotes the probability of a particle being in the energy state $E_\alpha$, $T$ represents the absolute temperature, $k$ is the Boltzmann constant, and $Z$, known as the partition function, normalizes the probabilities over all possible states of the system. The partition function is computed as the sum of the Boltzmann factors for all states, as shown in Equation \eqref{keq2}.

\begin{align}\label{keq2}
Z=\sum_\alpha exp(-\frac{E_\alpha}{kT})
\end{align}

The Boltzmann distribution is essential in statistical physics to understand the behavior of systems in thermal equilibrium.

\subsection{Probabilistic Modeling with Restricted Boltzmann Machine}
\begin{figure}
    \centering
    \includegraphics[width=0.9\linewidth]{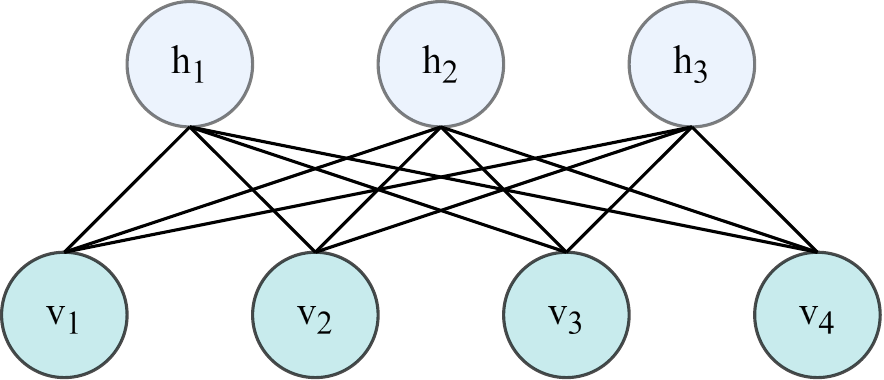}
    \caption{An RBM with 4 visible nodes and 3 hidden nodes.}
    \label{fig:An RBM with 4 visible nodes and 3 hidden nodes}
\end{figure}

In the landscape of machine learning, the Restricted Boltzmann Machine emerges as a trailblazer among unsupervised learning algorithms, laying a cornerstone for the construction of deep learning architectures. Functioning as an energy-based model, an RBM is uniquely characterized by its bipartite graph architecture, which comprises two distinct layers: the visible layer, directly interfacing with the input data, and the hidden layer, responsible for uncovering latent patterns and features. For instance, an RBM with 4 visible nodes and 3 hidden nodes can be visually represented as a bipartite graph $k_{4,3}$, as shown in Figure \ref{fig:An RBM with 4 visible nodes and 3 hidden nodes}. This architecture is distinctively characterized by the absence of intra-layer connections, confining node communication solely to the opposing layer. Such a design principle fosters fully interconnected inter-layer dynamics, crucial for the model's capacity to discern and encode the intricate patterns and interdependencies embedded in the data.

The operation of a Restricted Boltzmann Machine (RBM) is based on stochastic interactions between its visible and hidden layers, where the visible layer encodes input data and the hidden layer captures latent features. These interactions are governed by weights and biases, optimized to minimize the system's energy. Parameter learning is performed using the Contrastive Divergence (CD) algorithm, which iteratively adjusts the model to approximate the target data distribution.

In terms of network structure, an RBM is defined by:

\begin{enumerate}
\item{A vector $\boldsymbol{v} \in R^{K_v}$ in the visible layer, receiving and representing the input data, where ${K_v}$ denotes the length of the data.}
\item{A vector $\boldsymbol{h} \in R^{K_h}$ in the hidden layer, capturing the latent features and patterns within the data, where ${K_h}$ denotes the length of the feature.}
\item{A weight matrix $\boldsymbol{W} \in R^{K_v \times K_h}$, dictating the interaction strengths between units of the visible and hidden layers.}
\item{Biases $\boldsymbol{a} \in R^{K_v}$ and $\boldsymbol{b} \in R^{K_h}$, providing a base activation level for each unit.}
\end{enumerate}

The energy function $E(\boldsymbol{v},\boldsymbol{h})$ of RBM is given by Equation \eqref{eq3}. The joint probability distribution $p(\boldsymbol{v},\boldsymbol{h})$ is given by Equation \eqref{eq4}, where $Z$ defined by Equation \eqref{eqZ} is the partition function that ensures the probabilities sum to one.

In this work, RBM is harnessed to generate synthetic attack data for CAN. By generating new data samples conforming to the statistical traits of the original data, RBM addresses the data scarcity issue prevalent in CAN network security research.

\begin{equation}\label{eq3}
\begin{split}
E(\boldsymbol{v},\boldsymbol{h})&=-\sum_i {a_iv_i}-\sum_j {b_jh_j}-\sum_i {\sum_j {v_iw_{i,j}h_j}}\\
&=-\boldsymbol{a}^{\mathrm{T}} \boldsymbol{v}-\boldsymbol{b}^{\mathrm{T}} \boldsymbol{h}-\boldsymbol{v}^{\mathrm{T}} \boldsymbol{Wh}
\end{split}
\end{equation}

\begin{equation}\label{eq4}
\begin{aligned}
p(\boldsymbol{v},\boldsymbol{h})&=\frac{1}{Z}exp(-E(\boldsymbol{v},\boldsymbol{h}))\\
&=\frac{1}{Z}exp(\boldsymbol{a}^{\mathrm{T}} \boldsymbol{v})exp(\boldsymbol{b}^{\mathrm{T}} \boldsymbol{h})exp(\boldsymbol{v}^{\mathrm{T}} \boldsymbol{Wh})
\end{aligned}
\end{equation}

\begin{equation}\label{eqZ}
\begin{aligned}
Z=\sum_{\boldsymbol{v,h}} exp(-E(\boldsymbol{v},\boldsymbol{h}))
\end{aligned}
\end{equation}

\section{Methodology}
This section details our strategic methodology devised to confront the problem of dataset insufficiency and unbalanced sample distribution in CAN security research. We first elaborate on the CAN Data Processing Module, which is meticulously designed to convert raw CAN data into a format compatible with the Restricted Boltzmann Machine. Subsequently, the Negative Sample Generation Module capitalizes on the trained RBM to generate data that reflects the complexity of CAN traffic.

\subsection{CAN Data Processing Module}
In the domain of automotive cybersecurity, the authenticity of synthetic data is crucial for the effectiveness of Intrusion Detection Systems. Accordingly, we have meticulously devised a CAN Data Processing Module. This module is dedicated to converting raw CAN data into a format suitable for RBM training, guaranteeing that the input data for the RBM is not only uniform but also capable of embodying the intricate dynamics characteristic of CAN traffic. 

When a CAN message is transmitted across the CAN bus, three essential attribute fields come to light: the Identifier (ID), Data Length Code (DLC), and Data Content (DATA). In tandem with these is the Timestamp, which precisely pinpoints the bit time when the CAN controller receives the message, synchronized with the initiation time of the channel. Notably, while the Timestamp is not visibly presented within the message, the CAN controller can retrieve a 16-byte counter value from its CAN\underline{~}RDTxR register. Table \ref{tab:Extraction of Information Attributes from CAN Data Frames.} showcases the preliminary extraction outcomes of CAN message attributes for the aforementioned four distinct fields. Subsequently, the following data processing procedures are implemented in this research: 

\begin{enumerate}
\item{Exclusion of Non-Standard Data Frames: Our methodological approach commences with a stringent selection process, wherein we systematically eliminate any CAN data frames exhibiting a Data Length Code (DLC) that deviates from the standard length of 8. Given that the DLC for all CAN data frames within our purview is uniformly set to 8, this specific field demands no further examination in the course of our analysis.}
\item{Timestamp Normalization: We implement a timestamp normalization procedure to reduce variance in data size across different parts of CAN frames. Specifically, we conduct a differential operation on a per-frame basis. Subsequently, the resultant values undergo multiplication and are rounded to a positive integer in decimal notation. Such normalization guarantees the consistency of temporal data, rendering it optimally suitable for the ensuing training phase.}
\item{Conversion to Binary: Uniformly transforming the hexadecimal Identifier (ID) and Data Content (DATA) fields into their binary equivalents constitutes the final step in our data preprocessing procedures. Each byte of data within the ID and DATA fields is accurately converted from decimal values to a sequence of 8 binary bits. Simultaneously, the Timestamp field, which is represented in decimal, is converted into a 16-bit binary sequence. This binarization process renders the training data congruent with the model's inherent binary architecture.}
\end{enumerate}

Notably, for the DoS Attack dataset, we retain only the timestamp field to optimize the training process, recognizing that the flood data frames are marked by a data field of zeros and a high priority ID denoted by value of 0. This streamlined approach allows the RBM to focus on learning the probability distribution of the timestamp field, which is equivalent to understanding the entire data frame's distribution.

After completing these steps, the Timestamp field corresponds to 16 binary bits, the ID field corresponds to 16 binary bits and each byte of data in the Content field corresponds to 8 binary bits, totaling 64 bits. The processed data frames are then transformed into vectors suitable for RBM training. For general attack datasets, each CAN data frame is converted into a 96-dimensional vector, while for the DoS Attack dataset, a 16-dimensional vector (only Timestamp) is used. By meticulously processing the CAN data through these steps, we prepare a dataset that is not only reflective of the complex and variable nature of CAN traffic but also conducive to RBM training.

Taking the Spoofing GEAR Attack data as an illustration, the information attributes of the further processed data frames are presented in Table \ref{tab:Further Processed CAN Data.}.

\begin{table}
\centering
\caption{\label{tab:Extraction of Information Attributes from CAN Data Frames.}Extraction of Information Attributes from Raw CAN Data.}
\begin{tabular}{cccc}
   \toprule
   Timestamp & ID & DLC & Content \\
   \midrule
   1478195721.913135 & 0260 & 8 & fb, 7f, 00, 00, 40, 7f, 09, 22 \\
   1478195721.913370 & 02a0 & 8 & 00, 00, 00, 00, 0e, 23, 29, d0 \\
   1478195721.913607 & 0329 & 8 & 05, 21, 74, 09, 21, 21, 00, 6e \\
   1478195721.913844 & 0545 & 8 & fe, 5d, 00, 00, 00, 3c, 00, 00 \\
   1478195721.915732 & 02b0 & 8 & 19, 21, 21, 30, 08 \\
   1478195721.918460 & 0430 & 8 & 64, 00, 9b, 1d, 97, 02, bd, 00 \\
   1478195721.918689 & 04b1 & 8 & 0c, be, 7f, 14, 11, 20, 00, 14 \\
   \bottomrule
\end{tabular}
\end{table}

\begin{table}
\centering
\caption{\label{tab:Further Processed CAN Data.}Further Processed CAN Data from Spoofing GEAR Attack Data.}
\begin{tabular}{ccccccccccc}
   \toprule
   digit & 1bit & 2bit & 3bit & 4bit & ... & 94bit & 95bit & 96bit \\
   \midrule
    & 0 & 0 & 0 & 0 & ... & 0 & 1 & 0 \\
    & 0 & 0 & 0 & 0 & ... & 1 & 0 & 0 \\
    & 0 & 0 & 0 & 0 & ... & 0 & 0 & 0 \\
    & 0 & 0 & 0 & 1 & ... & 1 & 0 & 0 \\
    & 0 & 1 & 0 & 0 & ... & 1 & 1 & 1 \\
    & 0 & 0 & 0 & 0 & ... & 0 & 0 & 1 \\
    & 0 & 0 & 0 & 1 & ... & 1 & 0 & 1 \\
   \bottomrule
\end{tabular}
\end{table}

\subsection{Negative Sample Generation Module}
As a generative model, the Restricted Boltzmann Machine (RBM) aims to determine optimal network weights and bias values, with the ultimate goal of minimizing the system's energy state. This optimization empowers the RBM to generate data samples that highly approximate the statistical characteristics of the original dataset.

The training protocol for an RBM typically employs the Contrastive Divergence (CD) algorithm, which functions based on the principle of gradient ascent to iteratively enhance the model's parameters. The detailed procedural steps during the training process are outlined as follows:

\begin{enumerate}
\item{Parameter Initialization: The visible layer bias vector $\boldsymbol{a}$, the hidden layer bias vector $\boldsymbol{b}$ and the inter-layer weights matrix $\boldsymbol{W}$ are initialized.}
\item{Forward Propagation of the Visible Layer: After the visible layer receiving the input training data, given an input vector $\boldsymbol{v}$ from it, the activation probabilities for the hidden layer nodes are computed by Equation \eqref{eq5}, capturing the initial representation of the input data.}

\begin{align}\label{eq5}
p(h_i=1\mid\boldsymbol{v})=\sigma(b_i+\sum_{j=1}^n {w_{i,j}v_j})
\end{align}

\item{Hidden Layer Sampling: By leveraging the calculated activation probabilities, the state of the hidden layer is sampled. This process yields a compressed representation of the input data, distilling the essential features and facilitating further analysis.}

\item{Backward Propagation of the Hidden Layer: Subsequently, the sampled state of the hidden layer is employed to estimate the reconstruction probabilities for the nodes of the visible layer. The objective is to reconstruct the input data from the hidden layer's vantage point. The reconstruction probabilities of each node in the visible layer are computed in accordance with Equation \eqref{eq6}.}

\begin{align}\label{eq6}
p(v_i=1\mid\boldsymbol{h})=\sigma(a_i+\sum_{j=1}^n {w_{i,j}h_j})
\end{align}

\item{Visible Layer Sampling: Based on the aforementioned reconstruction probabilities, a sample of the visible layer is acquired. This step consummates the round-trip process that traverses from the visible layer to the hidden layer and then back to the visible layer, thereby facilitating the iterative refinement of the model's understanding of the input data.}
\item{Parameter Update: The Contrastive Divergence (CD) algorithm is utilized to modify the model parameters. By minimizing the disparity between the input and the reconstructed samples, the model's ability to accurately represent the data distribution is enhanced, facilitating a more precise portrayal of the underlying data characteristics.}
\item{Iteration: Steps 2 to 6 are iteratively repeated until the model reaches convergence or attains a preset number of iterations. This iterative process guarantees that the Restricted Boltzmann Machine (RBM) thoroughly grasps the underlying distribution of the data.}
\end{enumerate}

The specific steps of the learning process of the RBM for the target data distribution are clearly presented in Algorithm \ref{alg:rbm-cd}. 

\begin{algorithm}
\caption{RBM Training using Contrastive Divergence Algorithm}\label{alg:rbm-cd}
\begin{algorithmic}[1]

\Require Training data $\mathcal{V} = \{\boldsymbol{v}^1, \boldsymbol{v}^2, \ldots, \boldsymbol{v}^m\}$, learning rate $\eta$, number of hidden units $k$, number of iterations $T$

\State Initialize visible layer bias vector $\boldsymbol{a}$, hidden layer bias vector $\boldsymbol{b}$, and inter - layer weights matrix $\boldsymbol{W}$ randomly

\For{$t = 1$ to $T$}
    
    \For{each $\boldsymbol{v} \in \mathcal{V}$}

        \For{$i = 1$ to $k$}
            
            \State $p(h_i = 1|\boldsymbol{v})=\sigma(b_i+\sum_{j = 1}^n{w_{i,j}v_j})$ 
            
            \Comment{Equation \eqref{eq5}}
        
        \EndFor
        
        \State Sample hidden layer state $\boldsymbol{h}$ from $p(h_i = 1|\boldsymbol{v})$

        \For{$j = 1$ to $n$}
            
            \State $p(v_j = 1|\boldsymbol{h})=\sigma(a_j+\sum_{i = 1}^k{w_{i,j}h_i})$ 
            
            \Comment{Equation \eqref{eq6}}
        
        \EndFor
        
        \State Sample visible layer state $\boldsymbol{v}'$ from $p(v_j = 1|\boldsymbol{h})$

        \State $\Delta\boldsymbol{a}=\eta(\boldsymbol{v}-\boldsymbol{v}')$
        
        \State $\Delta\boldsymbol{b}=\eta\mathbb{E}_{p(h|\boldsymbol{v})}[\boldsymbol{h}]-\eta\mathbb{E}_{p(h|\boldsymbol{v}')}[\boldsymbol{h}]$
        
        \State $\Delta\boldsymbol{W}=\eta\boldsymbol{v}\boldsymbol{h}^T-\eta\boldsymbol{v}'\boldsymbol{h}^T$
        
        \State $\boldsymbol{a}=\boldsymbol{a}+\Delta\boldsymbol{a}$
        
        \State $\boldsymbol{b}=\boldsymbol{b}+\Delta\boldsymbol{b}$
        
        \State $\boldsymbol{W}=\boldsymbol{W}+\Delta\boldsymbol{W}$
    
    \EndFor

\EndFor

\State \Return $\boldsymbol{a}$, $\boldsymbol{b}$, $\boldsymbol{W}$

\end{algorithmic}
\end{algorithm}

In the above pseudocode:
- $\sigma(x)=\frac{1}{1 + e^{-x}}$ is the sigmoid function.
- $\mathbb{E}_{p(h|\boldsymbol{v})}[\boldsymbol{h}]$ represents the expected value of $\boldsymbol{h}$ under the distribution $p(h|\boldsymbol{v})$, and similar for $\mathbb{E}_{p(h|\boldsymbol{v}')}[\boldsymbol{h}]$.
- $n$ is the number of visible units and $m$ is the number of training samples.

After training, the Restricted Boltzmann Machine (RBM) acquires the capacity to generate novel data samples that emulate the statistical traits of the original dataset. According to Equations \eqref{eq5} and \eqref{eq6}, through a certain number of iterative transmissions between the visible layer and the hidden layer, the finally reconstructed vector conforms to the probability distribution of the original data.

\begin{figure}
    \centering
    \includegraphics[width=1.0\linewidth]{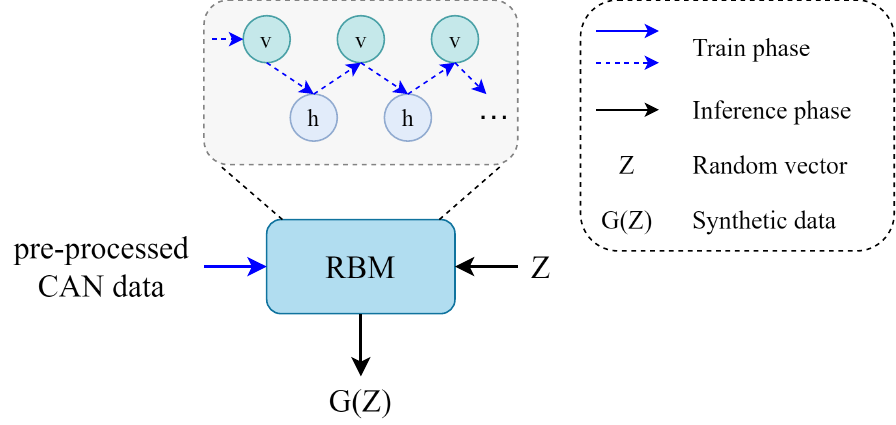}
    \caption{Demonstration of the Negative Sample Generation Module.}
    \label{fig:system stucture}
\end{figure}

The Negative Sample Generation Module based on RBM is shown in Figure \ref{fig:system stucture}. It is noteworthy that during the training stage, the training data processed by the CAN Data Processing Module is input from the visible layer. In contrast, during the inference stage, a random vector is input from the hidden layer and, after iteration, finally output from the visible layer.

\section{Experiments}
In this section, we introduce the experimental framework utilized to validate our RBM-based approach for generating synthetic CAN attack data. The experimental environment setup, dataset preparation, RBM model configuration, evaluation criteria, and results analysis are meticulously elaborated to showcase the effectiveness of our methodology in enhancing the performance of Intrusion Detection Systems.

\subsection{Environment}
In the experiments, the computational environment was configured with a 64-bit Intel(R) Core(TM) i5-13400F CPU operating at 2.5 GHz. The Python 3.11 programming language was employed to train and test our RBM model.

\subsection{Dataset}
In this study, we opted to employ the Car Hacking Dataset compiled by the Hacking and Countermeasure Research Lab (HCRL). The data samples within this dataset, captured via OBD-II port logging, have been meticulously classified into normal operational data, DoS Attack data, Fuzzy Attack data, Spoofing GEAR Attack data, and Spoofing RPM Attack data. The dataset meticulously records crucial CAN traffic parameters, such as timestamps, CAN IDs, DLCs, data bytes, and binary message flags (denoted as `T' for injected and `R' for regular), with detailed recordings in 30 - 40 minute intervals, each containing 300 instance injections.

\subsection{Configuration of the RBM Model Structure}
When utilized as a generative model, the Restricted Boltzmann Machine (RBM) iteratively optimizes the specific values of the inter-layer connection weights as well as the biases of both the visible and hidden layers during its training process. Nevertheless, ascertaining the number of nodes in each layer and setting the dimensions of the weight matrix necessitate manual investigation and calibration. In this research, in view of the four distinct types of cyber-intrusion data, we have devised the architecture of the Restricted Boltzmann Machine (RBM) as shown in Table \ref{tab:Specific Configuration of the RBM Structure.}.

\begin{table}
\centering
\caption{\label{tab:Specific Configuration of the RBM Structure.}Specific Configuration of the RBM Structure.}
\setlength\tabcolsep{3pt}
\begin{tabular}{cccc}
   \toprule
   Attack Type & Data Length & Visible Layer & Hidden Layer \\
   \midrule
   Dos & 16 bits & 16 nodes & 8 nodes \\
   Fuzzy & 96 bits & 96 nodes & 32 nodes \\
   Spoofing GEAR & 96 bits & 96 nodes & 32 nodes \\
   Spoofing RPM & 96 bits & 96 nodes & 32 nodes \\
   \bottomrule
\end{tabular}
\end{table}

\subsection{Evaluation Metrics}
The validation of the synthetic CAN network attack data generated by the Restricted Boltzmann Machine (RBM) is carried out via two principal approaches: similarity analysis in comparison with the original dataset and performance evaluation of the Intrusion Detection System (IDS) following data augmentation.

\subsubsection{Similarity Analysis}
The fidelity of the synthetic data is evaluated by contrasting it with the genuine network intrusion data procured from real-world scenarios. This is accomplished through the computation of the Cosine Similarity and the Pearson Correlation Coefficient between the generated samples and the original data samples. Elevated values of these similarity metrics signify that the synthetic data closely resemble the statistical characteristics of the original dataset. Equation \eqref{eq7} denotes the Cosine Similarity, while Equation \eqref{eq8} denotes the Pearson Correlation Coefficient, where X and Y stand for the data vectors being compared.

\begin{align}\label{eq7}
s=\frac{X{\ast}Y}{\Vert X \Vert\Vert Y \Vert}=\frac{\sum_{i=1}^n {X_i{\times}Y_i}}{\sqrt{\sum_{i=1}^n ({{X_i})^2}}{\times}\sqrt{\sum_{i=1}^n ({{Y_i})^2}}}
\end{align}

\begin{align}\label{eq8}
r=\frac{\sum_{i=1}^n {(X_i-\Bar{X}){\times}(Y_i-\Bar{Y})}}{\sqrt{\sum_{i=1}^n ({{X_i}-\Bar{X})^2}}{\times}\sqrt{\sum_{i=1}^n ({{Y_i}-\Bar{Y})^2}}}
\end{align}

\subsubsection{Performance Evaluation of IDS}
The effectiveness of the synthetic data is further corroborated by assessing the performance of two intrusion detection models, namely CANet \cite{Zhang2019CANetCS} and EfficientNet \cite{Tan2019EfficientNetRM}, prior to and subsequent to augmenting the training dataset with the samples generated by the RBM.

\begin{itemize}
\item{CANet, designed for few - shot semantic segmentation, shows excellent performance. It consists of a two - branch dense comparison module for multi - level feature comparison and an iterative optimization module for refining results. An attention mechanism is used in the k - shot setting. These components work together to achieve state - of - the - art results in few - shot segmentation tasks. }
\item{EfficientNet, designed as an image classification network capable of achieving high accuracy with relatively few parameters and FLOPS, demonstrates remarkable performance. It consists of a baseline network structure based on mobile inverted bottleneck MBConv with squeeze-and-excitation optimization and a compound scaling mechanism. The compound scaling method effectively balances network width, depth, and resolution for enhanced performance.}
\end{itemize}

To assess the effectiveness of the intrusion detection models with respect to dataset augmentation, a conventional set of classification metrics has been employed.

Accuracy, Recall, and Precision, which are derived from the counts of true negatives (TN), true positives (TP), false negatives (FN), and false positives (FP), are defined in the following manner:

\begin{align}\label{eq9}
Accuracy=\frac{TP+TN}{TP+TN+FP+FN}
\end{align}

\begin{align}\label{eq10}
Recall=\frac{TP}{TP+FN}
\end{align}

\begin{align}\label{eq11}
Precision=\frac{TP}{TP+FP}
\end{align}

Moreover, the F1-score serves as a highly informative metric in cases of unbalanced class distribution, providing a more refined assessment compared to accuracy solely. It is computed from the harmonic mean of precision and recall, as presented below:

\begin{align}\label{eq12}
F1=\frac{2{\times}precision{\times}recall}{precision+recall}
\end{align}

The Area Under the Curve (AUC) represents a crucial metric for assessing the effectiveness of classification models, especially in the context of binary classification scenarios. AUC captures the model's discriminative ability by characterizing the relationship between the true positive rate (TPR) and the false positive rate (FPR) over a range of classification thresholds.

\subsection{Experimental Results}

\subsubsection{Analysis of Similarity Results}
Take Spoofing GEAR Attack CAN data for instance, as observed from Table \ref{tab:G}, the generated dataset's rows 1, 2, 4, 6, and 7 consistently exhibit an ID value of 1087. This consistency implies a scenario where an adversary has compromised a specific ECU node (ID 1087), assuming control to disseminate deceptive data frames with the specific ID, thereby disrupting normal network operations.

\begin{table}
\centering
\caption{\label{tab:G}Generated Spoofing GEAR Attack data samples.}
\begin{tabular}{cccc}
   \toprule
   Timestamp & ID & DLC & Content \\
   \midrule
   5808 & 1087 & 8 & 1, 69, 96, 255, 107, 0, 0, 16 \\
   18031 & 1087 & 8 & 1, 4, 96, 127, 99, 4, 0, 0 \\
   1182 & 63 & 8 & 1, 69, 96, 255, 107, 0, 0, 8 \\
   9690 & 1087 & 8 & 1, 69, 396, 255, 107, 0, 0, 0 \\
   4988 & 1055 & 8 & 1, 69, 96, 255, 107, 0, 36, 0 \\
   34204 & 1087 & 8 & 1, 69, 96, 255, 107, 0, 0, 0 \\
   34038 & 1087 & 8 & 1, 69, 97, 255, 111, 0, 0, 0 \\
   \bottomrule
\end{tabular}
\end{table}

A comparison between the synthetic data and a subset of the original dataset shows a remarkable similarity in statistical properties. Table IV indicates that the average Cosine Similarity between the datasets is 0.8932 and the Pearson Correlation Coefficient is 0.8411. These metrics highlight the effectiveness of the synthetic data in reproducing the original attack patterns.

\begin{table}
\centering
\caption{\label{tab:Similarity Calculation Results.}Similarity Calculation Results.}
\begin{tabular}{cccc}
   \toprule
   Similarity Category & Average Value \\
   \midrule
   Cosine Similarity & 0.8932 \\
   Pearson Correlation Coefficient & 0.8411 \\
   \bottomrule
\end{tabular}
\end{table}

\subsubsection{Optimization of Intrusion Detection Based on Generated Data Samples}
Our method involves transforming CAN data frame sets into a visual form by integrating each set of 27 data frames into one image, thereby conforming to the data structures required by CANet and EfficientNet models. The Car Hacking Dataset is strategically categorized into different groups as follows:

\begin{itemize}
\item{The normal dataset is labeled as ``normal".}
\item{The DoS Attack dataset is labeled as ``dos".}
\item{The Fuzzy Attack dataset is labeled as ``fuzzy".}
\item{The Spoofing Gear Attack dataset is labeled as ``gear".}
\item{The Spoofing RPM Attack dataset is labeled as ``rpm".}
\end{itemize}

The initial datasets, once transformed into a visual array, are precisely segmented based on their binary labels. This leads to two main categories: normal and abnormal data. The abnormal data is further divided into sub-datasets, namely dos, fuzzy, gear, and rpm, thus creating a structured classification system for the original CAN data. After classification, the dataset composition is as follows: the normal dataset had 1,356 images, the dos dataset contained 1,490 images, the fuzzy dataset included 1,770 images, the gear dataset comprised 2,595 images, and the rpm dataset held 2,838 images.

Subsequently, 80\% of the normal data and 70\% of the abnormal data are selected for the training set to train the classification model. The remaining 20\% of the normal data and 20\% of the abnormal data are used as the validation set to avoid overfitting during training and balance the model's performance. Moreover, 10\% of the remaining abnormal data is set as the test set to evaluate the classification performance of the model.

In this work, the RBM model is employed to capture the probabilistic details of the CAN intrusion data and generate a large dataset containing 10,000 CAN data frame samples for each of the DoS, Fuzzy, Spoofing Gear, and Spoofing RPM attack data types. These data are then converted into an image-based representation to augment the training set of the intrusion detection model. During this process, the composition of the validation and test sets remains unchanged. The enlarged training set is then used to improve the model's learning ability.

Table \ref{tab:Changes in the Accuracy.} illustrates that the accuracy of the CANet model significantly increased from 0.6477 to 0.9725 after data augmentation. Likewise, although the EfficientNet model had a low initial accuracy, it also had a remarkable improvement from 0.1067 to 0.1555. This suggests that the data generated by RBM effectively enhanced the feature space of the models, facilitating their ability to distinguish between normal and attack patterns in CAN traffic. 

\begin{figure}
    \centering
    \includegraphics[width=1.0\linewidth]{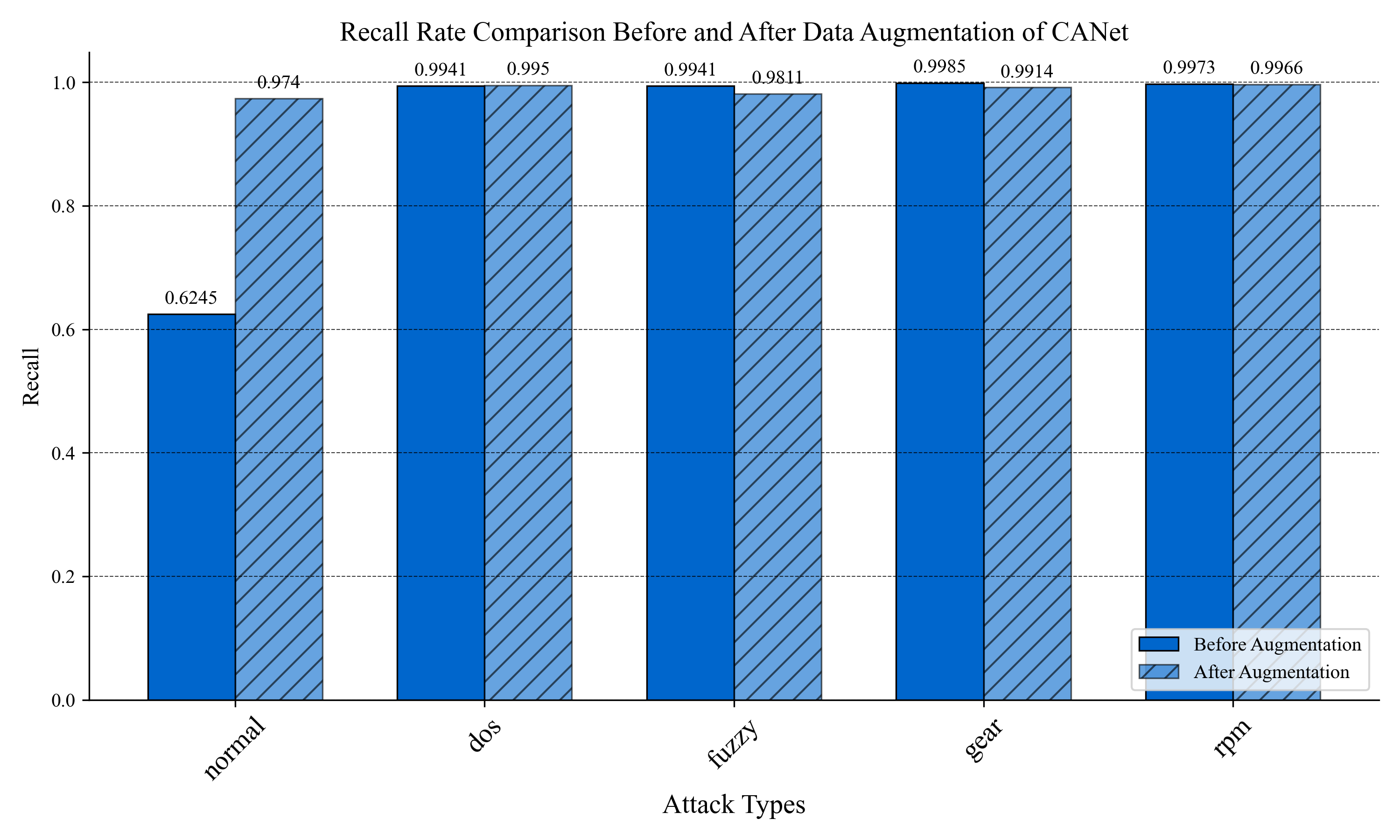}
    \caption{Recall Comparison of CANet.}
    \label{fig:CANet_recall_comparison}
\end{figure}

\begin{figure}
    \centering
    \includegraphics[width=1.0\linewidth]{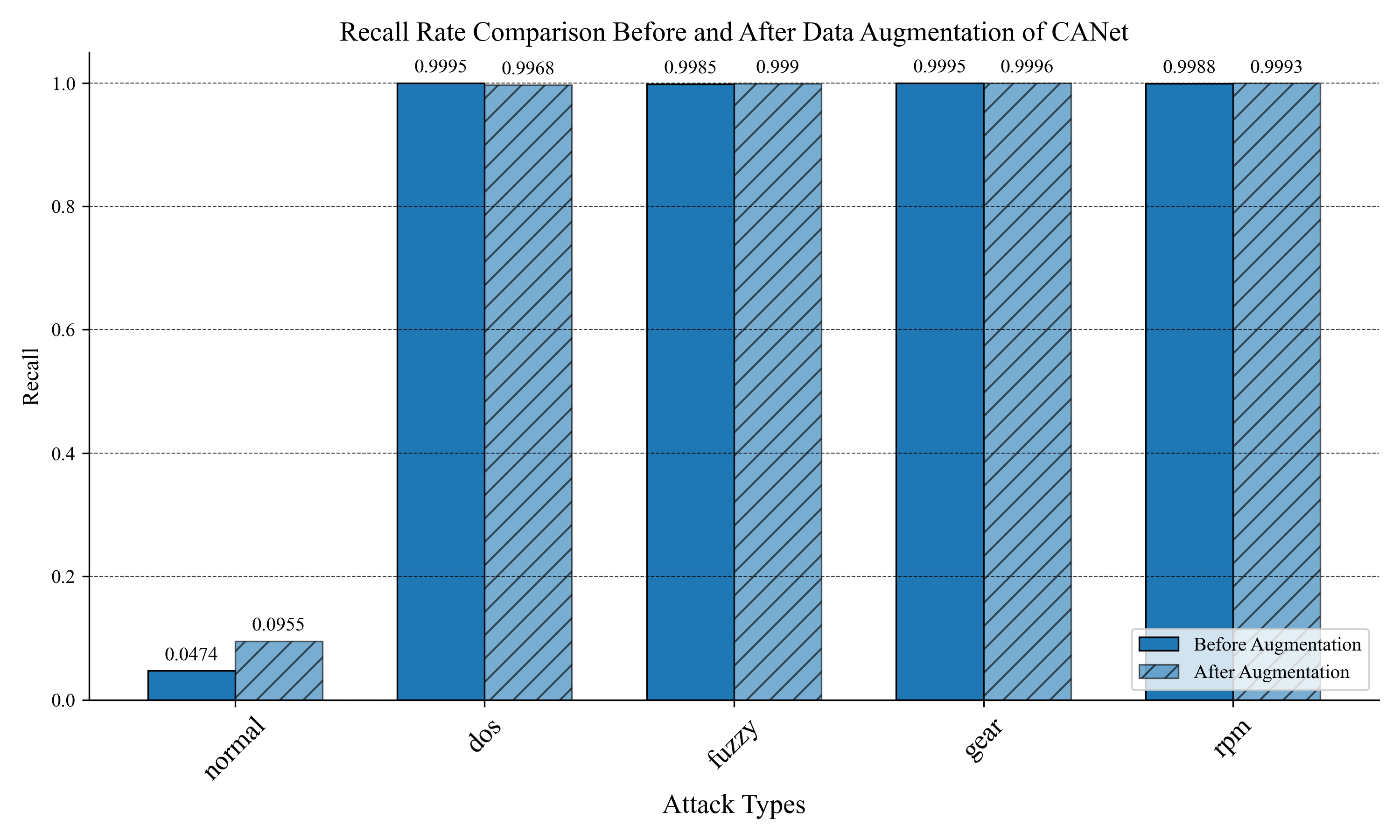}
    \caption{Recall Comparison of EfficientNet.}
    \label{fig:EfficientNet_recall_comparison}
\end{figure}

The recall rates, as shown in Figure \ref{fig:CANet_recall_comparison}, demonstrate a significant increase in the normal data category of the CANet model, rising from 0.6245 to 0.9740. Although there is a slight decrease in the fuzzy, gear, and rpm categories, they still remain at a very high level. For the EfficientNet model, as presented in Figure \ref{fig:EfficientNet_recall_comparison}, the recall rates also improve, despite a negligible decline in the dos attacks category. These results highlight the enhanced ability of the models to detect a broader range of attack instances while ensuring the detection of normal patterns.

\begin{figure}
    \centering
    \includegraphics[width=1.0\linewidth]{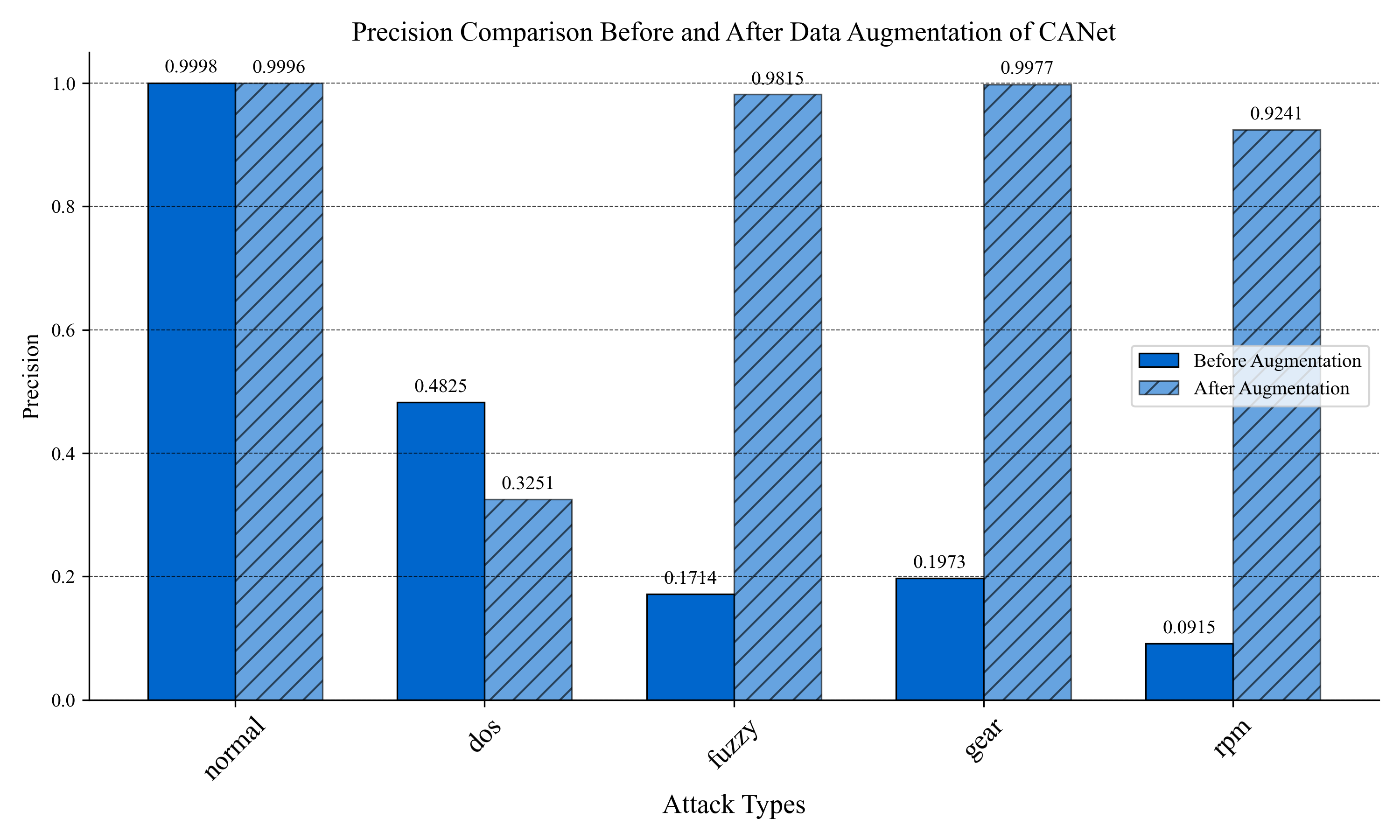}
    \caption{Precision Comparison of CANet.}
    \label{fig:CANet_precision_comparison}
\end{figure}

\begin{figure}
    \centering
    \includegraphics[width=1.0\linewidth]{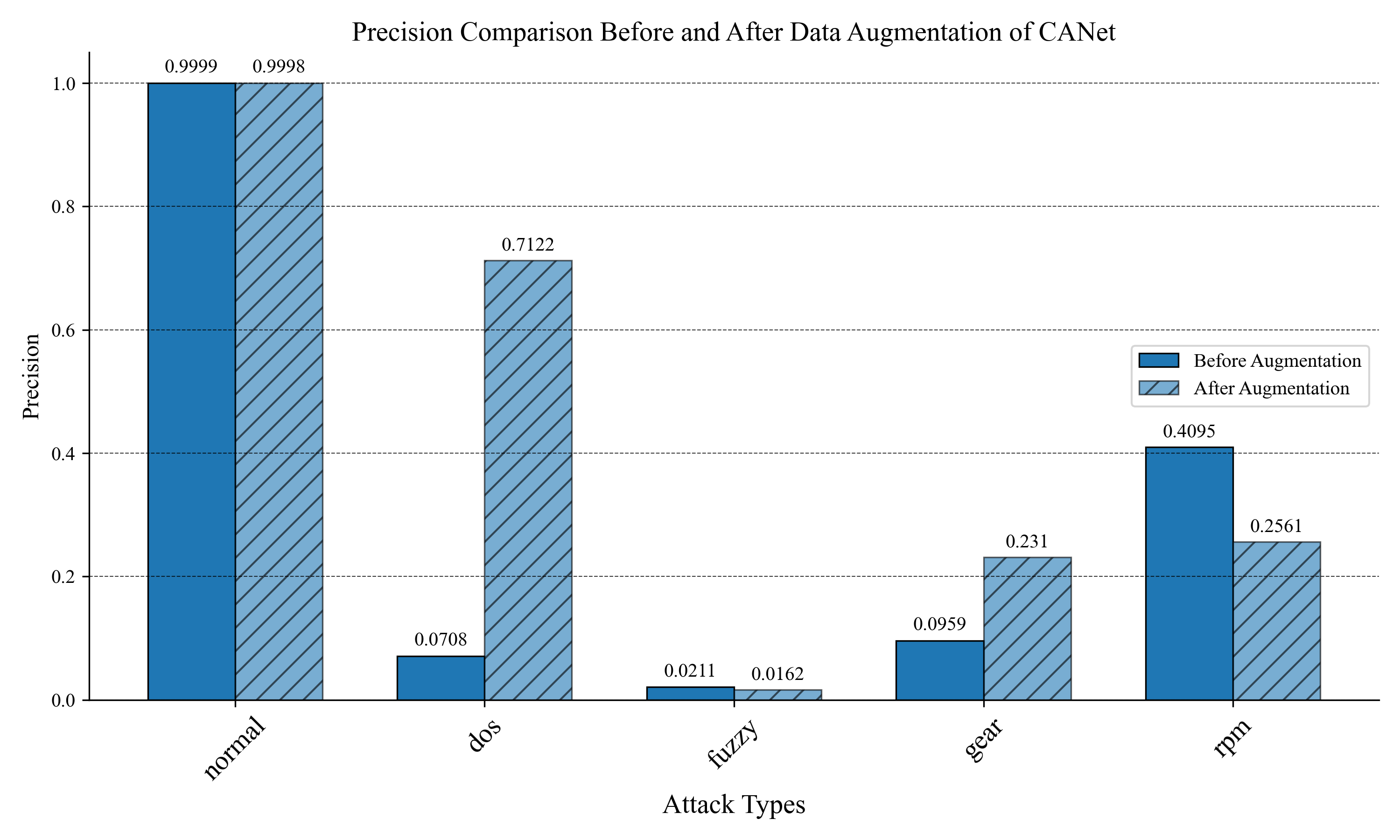}
    \caption{Precision Comparison of EfficientNet.}
    \label{fig:EfficientNet_precision_comparison}
\end{figure}

Precision is crucial for systems as false positives can trigger unnecessary responses. As depicted in Figure \ref{fig:CANet_precision_comparison} and \ref{fig:EfficientNet_precision_comparison}, the trends of precision are mixed but positive. For the CANet model, due to the increased recall, the precision of normal data decreased as expected. However, the precision of attack detection improved significantly. Meanwhile, for normal data, the precision of the EfficientNet model remains high and stable. In terms of attack types, the precision for DoS, Gear, and RPM attacks increases substantially after augmentation, indicating that the augmented data helps EfficientNet to better identify these attacks.

\begin{table}
\centering
\caption{\label{tab:Changes in the Accuracy.}Changes in the Accuracy.}
\begin{tabular}{ccc}
   \toprule
   Network & Accuracy & Value \\
   \midrule
   \multirow{2}*{CANet} & before augmentation & 0.6477 \\
    & after augmentation & 0.9725 \\
   \midrule
   \multirow{2}*{EfficientNet} & before augmentation & 0.1067 \\
    & after augmentation & 0.1555 \\
   \bottomrule
\end{tabular}
\end{table}

\begin{table}
\centering
\caption{\label{tab:Changes in the AUC.}Changes in the AUC.}
\begin{tabular}{ccc}
   \toprule
   Network & AUC & Value \\
   \midrule
   \multirow{2}*{CANet} & before augmentation & 0.9489 \\
    & after augmentation & 0.9564 \\
   \midrule
   \multirow{2}*{EfficientNet} & before augmentation & 0.9735 \\
    & after augmentation & 0.9562 \\ 
   \bottomrule
\end{tabular}
\end{table}

\begin{figure}
    \centering
    \includegraphics[width=1.0\linewidth]{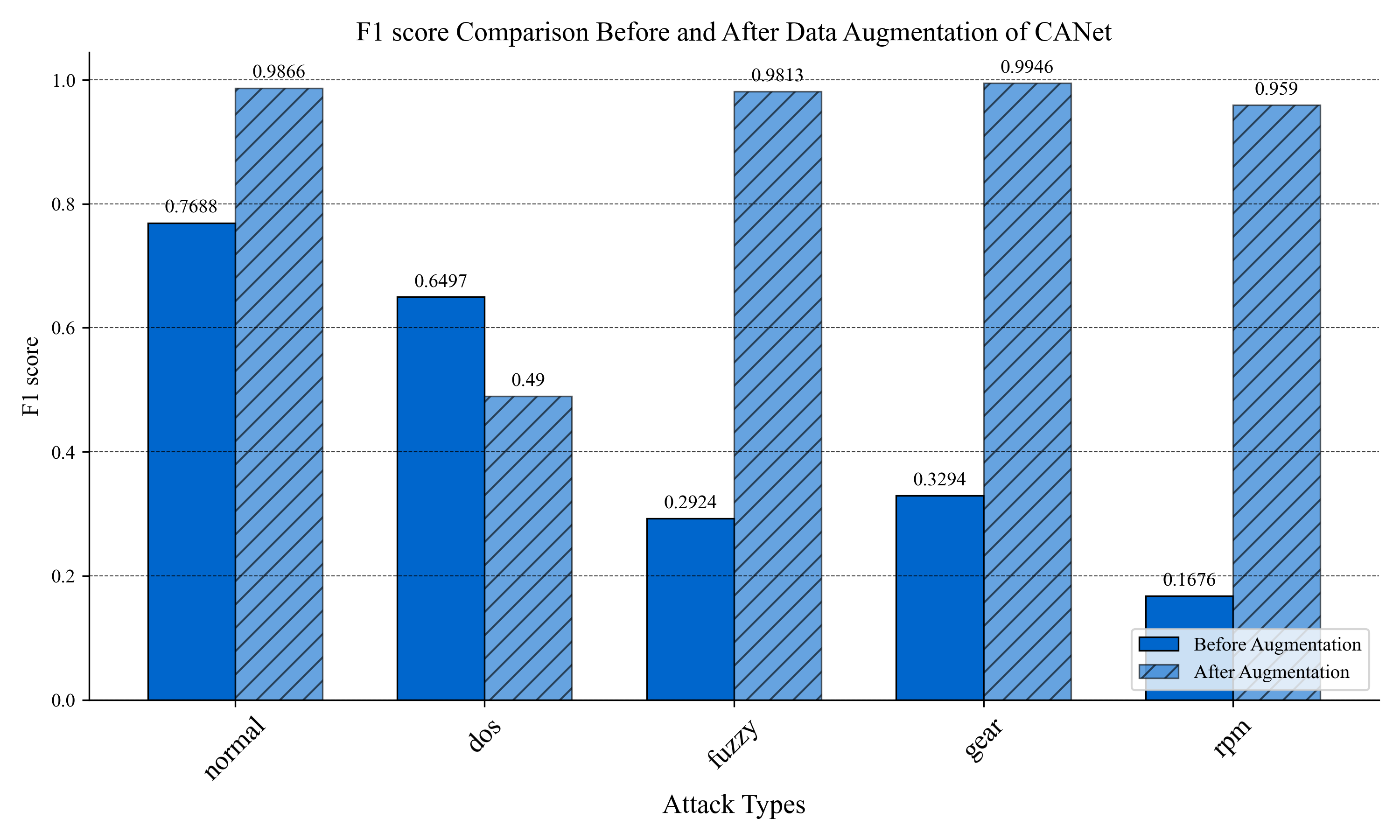}
    \caption{F1-score Comparison of CANet.}
    \label{fig:CANet_f1score_comparison}
\end{figure}

\begin{figure}
    \centering
    \includegraphics[width=1.0\linewidth]{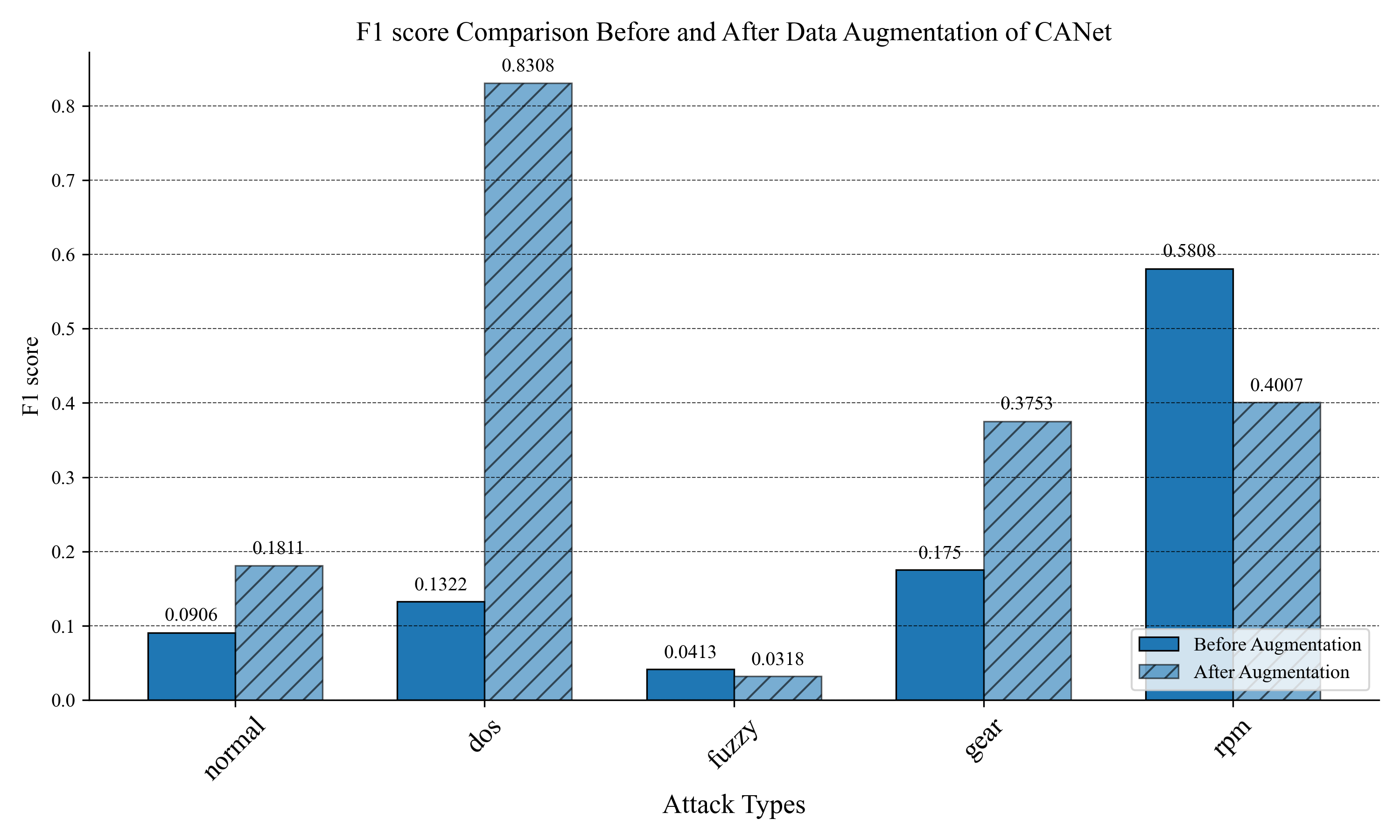}
    \caption{F1-score Comparison of EfficientNet.}
    \label{fig:EfficientNet_f1score_comparison}
\end{figure}

The F1 scores, which balance precision and recall, are depicted in Figure \ref{fig:CANet_f1score_comparison} and \ref{fig:EfficientNet_f1score_comparison}, and they demonstrate the overall enhancement of the models' performance. For the CANet model, the F1 scores increased significantly across most categories, except for DoS attacks, where there was a slight decline. For the EfficientNet model, its F1 - score has significantly increased for the normal, dos, and gear categories.

Finally, a detailed analysis of the AUC values in Table \ref{tab:Changes in the AUC.} offers a comprehensive evaluation of the models' discrimination abilities among different classes. The CANet model shows a significant improvement in its AUC value, increasing from 0.9489 to 0.9564. Although the EfficientNet model starts with a higher AUC value of 0.9735 but slightly decreases to 0.9562, this unexpected finding provides valuable information for future research. However, it should be stressed that despite this slight decline, the EfficientNet model's AUC value is still very high, indicating that our method has great potential to improve the performance of complex models such as EfficientNet.

In conclusion, using RBM - generated data to augment training datasets significantly improves the intrusion detection capabilities of both models. It also offers a robust framework for detecting a wider range of cyber - attacks in CAN systems. The results confirm the effectiveness of RBM for data generation, especially when real attack data is limited, contributing to the proactive cybersecurity measures for the automotive industry's digital transformation.

Our comprehensive method not only addresses data scarcity and imbalance issues but also supports the development of preemptive cybersecurity measures, keeping pace with the automotive industry's digital progress.

\subsection{Analysis of the Efficiency of the Negative Sample Generation Module}
The Negative Sample Generation Module proposed by us is a lightweight intrusion data generation network for the Internet of Vehicles, featuring an extremely high convergence speed. Figure 1 illustrates the convergence process of the loss function, that is, the reconstruction error of the visual layer, when the generation module learns the data distribution of SPOOFING GEAR ATTACK data. It can be observed that when the model is trained for 10 epochs, the network parameters have approached the optimal results. 

\begin{figure}
    \centering
    \includegraphics[width=1.0\linewidth]{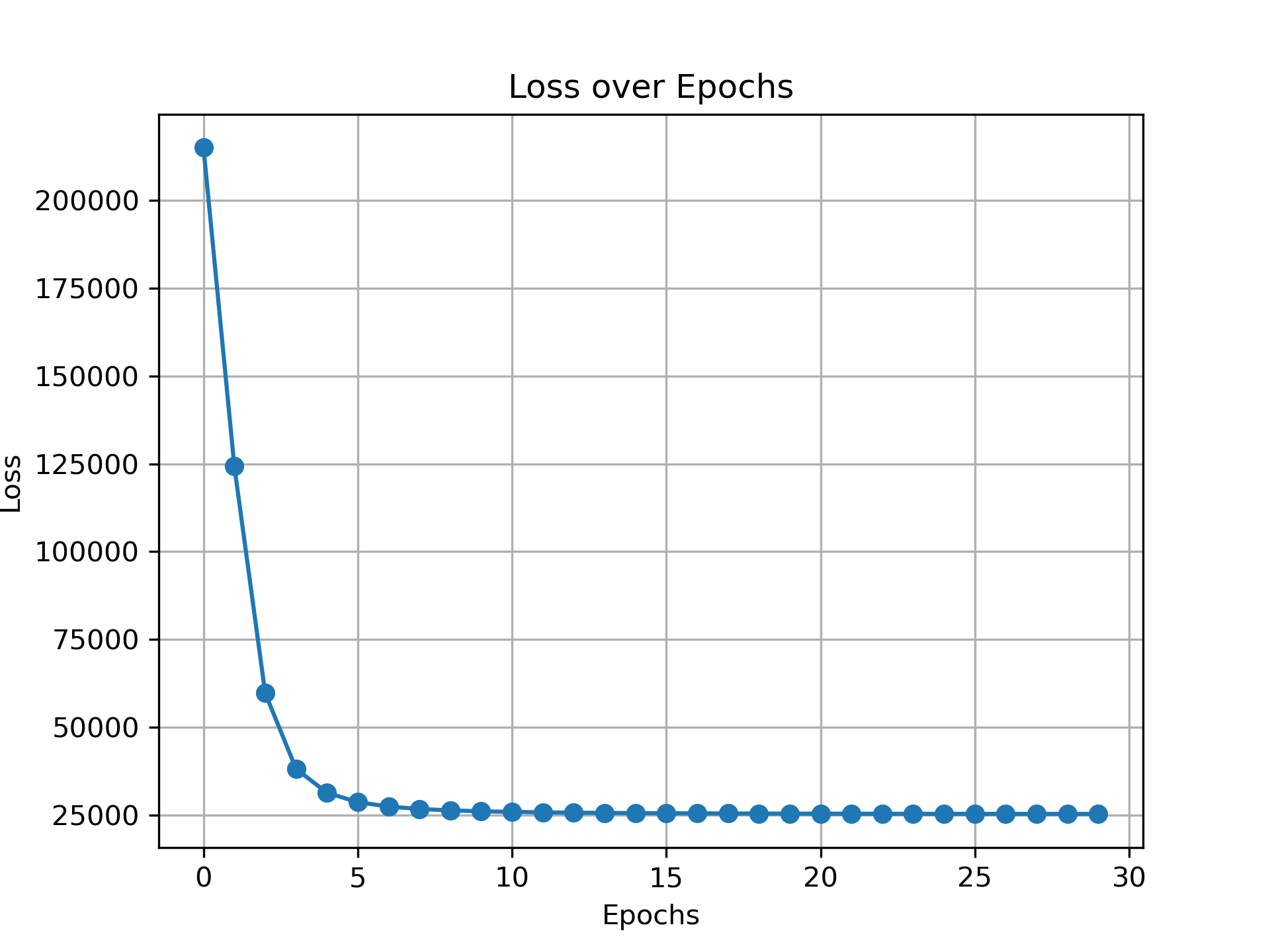}
    \caption{Convergence of RBM - based Generator}
    \label{fig:loss}
\end{figure}

It is worth noting that our Negative Sample Generation Module features a highly lightweight network structure. The visible layer contains 96 nodes, which are the biases of the visible layer. The hidden layer has 32 nodes, and there is a connection weight matrix of the shape \(96\times32\) between the two layers. The total number of parameters is \(96 + 32+96\times32=3200\) parameters.

\section{Discussion and Future Work}
The utilization of RBM-generated synthetic data has remarkably enhanced the detection capabilities of CAN-based IDS, as manifested by the notable improvements in multiple evaluation metrics of CANet and EfficientNet models. This approach effectively tackles the data scarcity issue in CAN network security research and offers a more proactive cybersecurity solution for the automotive industry.

However, several aspects demand further exploration. Firstly, the current manual adjustment of RBM parameters can be automated through advanced hyperparameter optimization techniques. This would not only streamline the model training process but also potentially enhance the model's adaptability and performance. For example, algorithms like grid search, random search, or more sophisticated Bayesian optimization methods could be employed to systematically explore the parameter space and identify the optimal configuration.

Secondly, investigating alternative RBM architectures and integrating other generative models holds great promise. Different RBM structures might capture more intricate data patterns and relationships, while combining multiple generative models could enrich the diversity of generated attack patterns. For instance, exploring deep belief networks (DBNs) which are composed of multiple stacked RBMs or incorporating variational autoencoders (VAEs) alongside RBMs could potentially lead to more comprehensive and realistic synthetic data.

Thirdly, the potential of real-time data generation and its influence on IDS performance remains an uncharted territory. Real-time generation of synthetic data could provide a dynamic training environment that enables IDS to promptly respond to emerging threats. This requires the development of efficient algorithms and systems capable of generating data on-the-fly without sacrificing quality or introducing significant computational overhead.

Finally, when deploying such systems in operational networks, a comprehensive security assessment is essential. This involves evaluating potential vulnerabilities introduced by the synthetic data generation process and ensuring the integrity and reliability of the network. Thorough testing and validation procedures need to be established to guarantee that the system functions as intended and does not pose any additional risks.

Future research efforts should focus on these directions to fully exploit the potential of RBMs and other related techniques in enhancing CAN network security and promoting the development of proactive cybersecurity measures in the automotive industry.

\section{Conclusion}
In this research, we addressed the critical issue of data scarcity and imbalance in CAN network security by developing an innovative methodology based on the Restricted Boltzmann Machine (RBM). Through the design of the CAN Data Processing Module and the Negative Sample Generation Module, we successfully transformed raw CAN data into a suitable format for RBM training and generated synthetic attack data that closely resembled the statistical properties of real CAN traffic.

Experimental results clearly demonstrated the effectiveness of our approach. The accuracy of CANet increased from 0.6477 to 0.9725, and that of EfficientNet improved from 0.1067 to 0.1555 after data augmentation. Other metrics such as recall, precision, F1-score, and AUC also showed significant enhancements, indicating that the generated data effectively enhanced the models' ability to distinguish between normal and attack patterns.

Our work not only contributes to the improvement of intrusion detection capabilities in CAN systems but also provides a promising solution for the automotive industry's cybersecurity challenges. By generating synthetic data, we overcame the limitations of insufficient attack data and offered a more proactive approach to network security.

In summary, this study represents a significant step forward in CAN network security research. The proposed RBM-based method has proven its value in enhancing the performance of IDS and promoting the development of proactive cybersecurity measures in the automotive industry.

\bibliographystyle{IEEEtran}
\bibliography{CANDataSynthetic}

%\printbibliography

\end{document}